\documentclass[12pt]{iopart}
\usepackage{iopams,setstack,amssymb,graphicx,multirow,hyperref,doi}

\hypersetup{
    colorlinks=true,       
    linkcolor=blue,          
    citecolor=blue,        
    filecolor=blue,      
    urlcolor=blue           
}

\graphicspath{{./}{./figs/}}

\begin{document}

\title{A constant pressure flowmeter for extreme-high vacuum}
\author{S. Eckel$^1$, D. S. Barker$^1$, J. Fedchak$^1$, E. Newsome$^1$, J. Scherschligt$^1$, R. Vest$^1$}
\address{$^1$Sensor Sciences Division, National Institute of Standards and Technology, Gaithersburg, MD 20899, USA}
\ead{stephen.eckel@nist.gov}

\begin{abstract}
We demonstrate operation of a constant-pressure flowmeter capable of generating and accurately measuring flows as low as $2\times10^{-13}$~mol/s.
Generation of such small flows is accomplished by using a small conductance element with $C\approx 50$~nL/s.
Accurate measurement then requires both low outgassing materials ($<1\times 10^{-15}$~mol/s) and small volume changes ($\approx 70$~$\mu$L).
We outline the present flowmeter's construction, detail its operation, and quantify its uncertainty.
The type-B uncertainty is $<0.2$~\% ($k=1$) over the entire operating range.
In particular, we present an analysis of its hydraulic system, and quantify the shift and uncertainty due to the slightly compressible oil.
Finally, we compare our flowmeter against a NIST standard flowmeter, and find agreement to within 0.5~\% ($k=2$).
\end{abstract}
\maketitle

\section{Introduction}
\label{sec:intro}
Constant pressure flowmeters provide high-accuracy low-range flows for a host of metrological applications~\cite{McCulloh1987, Fedchak2012}, including orifice flow standards for calibrating pressure gauges and leak standards.
Here, we present a new, fully-automated flowmeter capable of producing flows between $2\times10^{-13}$~mol/s and $1\times10^{-8}$~mol/s.
Using a 100:1 flow splitter in combination with a 100~L/s orifice flow standard, this flow range corresponds to generated pressures between $2\times10^{-11}$~Pa and $1\times10^{-5}$~Pa, extending NIST's flowmeters from the ultra-high vacuum (UHV) to the extreme-high vacuum (XHV).
Moreover, this flowmeter, combined with such an orifice flow standard, can directly measure the thermalized cross sections between various background gases and trapped atoms, enabling the comparison of both the cold atom vacuum standard (CAVS)~\cite{Scherschligt2017} and its portable counterpart (p-CAVS)~\cite{Eckel2018}.

Constant pressure flowmeters both generate and measure the flow by allowing gas to flow at rate $\dot{n}$ through a small constriction from a variable volume.
To measure $\dot{n}$, the volume is reduced by a feedback-controlled mechanism in order to keep its pressure constant.
The flow out of the constriction is then equal to the negative of the change in moles of the gas inside the variable volume.
Assuming an ideal gas at a constant temperature $T$ and pressure $p$, the flow equation is
\begin{equation}
    \label{eq:basic_measurement_equation}
    \dot{n} = -\frac{p}{R T}\frac{d V}{dt} + \dot{n}_{\rm OG}
\end{equation}
where $R$ is the molar gas constant, $V$ is the volume, $t$ is time, and $\dot{n}_{\rm OG}$ is any outgassing of the volume walls.  Note that $dV/dt<0$ and $\dot{n}_{\rm OG}>0$.

\begin{figure}
    \centering
    \includegraphics{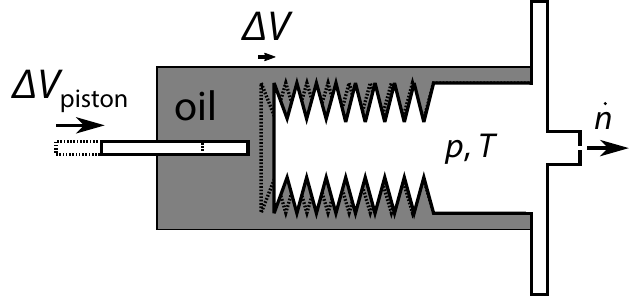}
    \caption{
    Basic idea of a hydraulic constant pressure flowmeter.
    A gas at pressure $p$ and temperature $T$ leaks out of a variable volume at a rate $\dot{n}$.
    To measure $\dot{n}$, a piston is inserted into oil immersing the variable volume, causing the latter to contract.
    For incompressible oil, the volume change of the piston is equal to that of the variable volume, i.e., $\Delta V_{\rm piston} = -\Delta V$.
    A PID changes the volume $\Delta V$ to keep $p$ constant; the rate of change of $\Delta V$ measures $\dot{n}$ through Eq.~\ref{eq:basic_measurement_equation}.}
    \label{fig:explanation}
\end{figure}

First-generation constant-pressure flowmeters at NIST changed the volume by plunging a piston directly into the variable volume.
This technique, while simple, suffered from large systematic errors due to leakage through the seal between the piston and the wall, restricting it to flows greater than $1\times10^{-10}$~mol/s.
For lower flows, a second flowmeter with the hydraulic system depicted in Fig.~\ref{fig:explanation} was used.
This design is the one we adopt for the present apparatus.
It encloses a flexible vacuum component, here a welded bellows, inside a closed container filled with incompressible oil.
To effect a change in volume, a precisely-dimensioned piston is inserted into the oil, which in turn exerts a force on the flexible volume component.
Assuming incompressible oil, the change of volume of the piston $\Delta V_{\rm piston}$ forces the volume of the variable volume to change by an equal amount, i.e. $\Delta V = -\Delta V_{\rm piston}$.
Outgassing of the metal walls of the variable volume generally restricted flows to greater than $1.0 \times 10^{-11}$~ mol/s for such legacy systems.
This systematic can also limit the uncertainty of the third type of constant pressure flowmeters, those that use a calibrated bellows rather than hydraulics~\cite{Jousten1993, Jousten1999, Jousten2002, Gronych2008, Berg2014}.

To measure flows with $\dot{n}\simeq 1.0 \times 10^{-13}$~mol/s with an accuracy of 0.2~\%, requires $\dot{n}_{\rm OG}<1.0 \times 10^{-15}$~mol/s, assuming that $\dot{n}_{\rm OG}$ can be quantified at the 10~\% level.
Outgassing in vacuum materials is predominantly from two processes, water desorption from surfaces and hydrogen diffusing from the bulk interior of stainless steel components.
Ordinarily, water desorption can be minimized by baking the system under vacuum at roughly 150 $^\circ$C for at least $24$ hours.
This procedure effectively strips away water from interior surfaces, leaving little water to desorb at ambient temperatures.
However, subsequent exposure to water can cause re-adsorption on the interior surfaces, increasing the water desorption rate.
Such exposure is possible, for example, from contamination in the gas used to fill the flowmeter, and thus a low-outgassing flowmeter must be designed for repeated bakes.
Hydrogen outgassing can be more difficult to minimize, requiring either a low hydrogen-outgassing material, like titanium or aluminium~\cite{Fedchak2021}, or a process of baking the vacuum components at higher temperatures~\cite{Mamun2013, Sefa2017, Fedchak2018}.

A common misconception is that the moving piston somehow drives the flow $\dot{n}$.
The piston merely {\it measures} $\dot{n}$; instead, $\dot{n}$ is produced through pressure difference across the constriction of conductance $C$.
The pressure $p$ is generally limited to $10\mbox{  Pa}<p<100\mbox{ kPa}$, set by the operational ranges of high-accuracy pressure transfer standards such as capacitance diaphragm gauges (CDGs) and resonant silicon gauges.
Assuming the flowmeter operates near 300~K, generation of flows $\dot{n} \simeq 1.0 \times 10^{-13}$~mol/s thus requires small conductances, $C<1$~$\mu$L/s.
Furthermore, under ideal operation $C = -dV/dt$, and thus smaller $C$ generally require smaller displacements.

\section{Theoretical background}
\label{sec:theory}
Most low gas-flow vacuum flowmeters, no matter the type, operate on a single principle.
Gas is allowed to leak from a sealed container, causing the number of moles contained $n$ to drop with time.
The output flow $\dot{n}_{\rm out}$ is then the negative of the rate of change of $n$, $\dot{n}$, plus the contribution of material outgassing $\dot{n}_{\rm OG}$.
Thus to determine the flow, one must measure $\dot{n}$.
Inserting the derivative of the ideal-gas law, $n=pV/RT$, one arrives at the ideal-gas measurement equation,
\begin{eqnarray}
    \dot{n}_{\rm out} & = &  -\frac{d}{dt}\left(\frac{p V}{RT}\right) + \dot{n}_{\rm OG} \\
    & = & - \frac{p\dot{V}}{R T} -\frac{\dot{p}V}{R T} +\frac{p V}{R T^2}\dot{T} + \dot{n}_{\rm OG}. \label{eq:full_measurement_equation}
\end{eqnarray}
The first three terms correspond to measurement of the rate of change one container or gas property to determine both $\dot{n}$ and $\dot{n}_{\rm out}$: (1) the rate of change of volume at constant pressure and temperature, i.e., the constant-pressure flowmeter; (2) the rate of change of pressure at constant volume and temperature, i.e., a constant-volume flowmeter; and (3) the rate of change of temperature at constant volume and pressure.
To our knowledge, the last type has never been demonstrated.\footnote{Indeed, constant-pressure and constant-volume flowmeters might be better described as constant-pressure-and-temperature and constant-volume-and-temperature flowmeters, respectively.}
No flowmeter is perfect, and thus Eq.~\ref{eq:full_measurement_equation} can be used to correct for any control errors.
For example, in a constant pressure flowmeter that experiences temperature and pressure drifts (i.e., $\dot{p},\dot{T}\neq 0$), Eq.~\ref{eq:full_measurement_equation} becomes the full measurement equation and captures those additional corrections.

\begin{figure}
    \centering
    \includegraphics{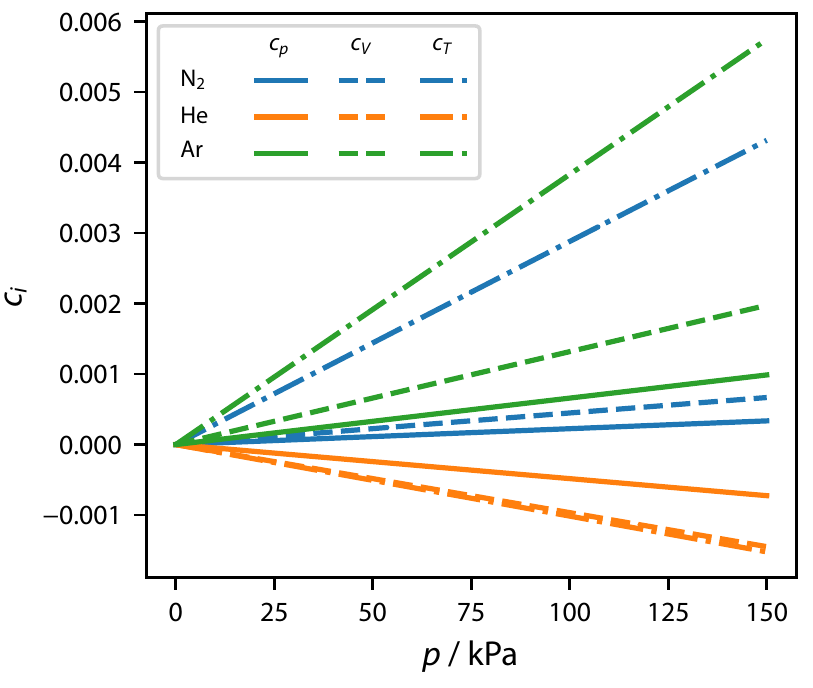}
    \caption{Non-ideal gas correction factors for N$_2$ (blue), He (orange), and Ar (green) at 295~K in constant-pressure flowmeters ($c_V$, solid), constant-volume flowmeters ($c_p$, dashed), and constant-volume-and-pressure flowmeters ($c_T$, dash-dotted).  Values of the virial coefficients come from the recommended values of Ref.~\cite{Rourke2021}.}
    \label{fig:non_ideal_gas}
\end{figure}
In addition, the gas in the volume being measured need not be ideal.
Non-ideal gases are commonly parameterized by the virial equation of state,
\begin{equation}
    \label{eq:virial_eos}
    \frac{p}{RT} = \frac{n}{V} + B_2(T) \left(\frac{n}{V}\right)^2 + B_3(T) \left(\frac{n}{V}\right)^3 + \cdots,
\end{equation}
where $B_i$ are the virial coefficients.
To quantify such corrections to lowest order, we differentiate Eq.~\ref{eq:virial_eos} with respect to $t$, solve for $\dot{n}$, expand small terms, group terms in a similar way to Eq.~\ref{eq:full_measurement_equation}, and insert the ideal gas solution for $n/V=p/RT$.
The resulting full measurement equation is
\begin{equation}
    \dot{n}_{\rm out} = - (\dot{n}_V + \dot{n}_p + \dot{n}_T) + \dot{n}_{\rm OG}, \label{eq:full_measurement_equation_with_virial}
\end{equation}
where
\begin{eqnarray}
    \dot{n}_V & \equiv & \frac{p\dot{V}}{R T}\left(1 + c_V\right) \\
    \dot{n}_p & \equiv & \frac{\dot{p}V}{R T}\left(1 + c_p\right) \\
    \dot{n}_T & \equiv & -\frac{p V}{R T^2}\dot{T}\left(1 + c_T\right)
\end{eqnarray}
with the virial corrections (up to second order in $p/R T$)
\begin{eqnarray}
    c_V & \equiv & -B_2 \frac{p}{RT} - B_3 \left(\frac{p}{RT}\right)^2\\
    c_p & \equiv & -2 B_2 \frac{p}{RT} - (3 B_3 - 4 B_2^2)\left(\frac{p}{RT}\right)^2\\
    c_T & \equiv & \left(T \frac{dB_2}{dT} - 2 B_2 \right) \frac{p}{RT} + \left[T\frac{dB_3}{dT} - (3 B_3 - 4 B_2^2) - 2 T B_2\frac{dB_2}{dT}\right]\left(\frac{p}{RT}\right)^2.\label{eq:non_ideal_T_correction}
\end{eqnarray}
Fig.~\ref{fig:non_ideal_gas} shows these corrections plotted as a function of pressure at 295~K.
These corrections do not depend on $\dot{n}$, the measured $\dot{p}$, $\dot{V}$, or $\dot{T}$ quantities, or the volume $V$, but only the mean pressure $p$ and $T$.
For $c_V$, the correction is due to the difference in the number of moles in the gas compared to ideal, i.e., $B_2 N/V\approx B_2 p/k_B T$.
For $c_p$, the factor of 2 in the first order term comes from the first derivative of that difference as the pressure changes in the gas.
Finally, for $c_T$, the first-order correction is comprised of two components: the change in the virial coefficient with temperature, $T(dB/dT)$, which is dominant for N$_2$, and the derivative of the correction of the number of moles in the gas from ideal as the temperature changes ($-2 B_2$).
For the constant-pressure flowmeter discussed here, the correction at a fill pressure of 100~kPa for N$_2$ is approximately 0.02~\%, only about a factor of 5 lower than our target uncertainty of 0.1~\%.

\section{Component quantification}
\begin{figure}
    \centering
    \includegraphics{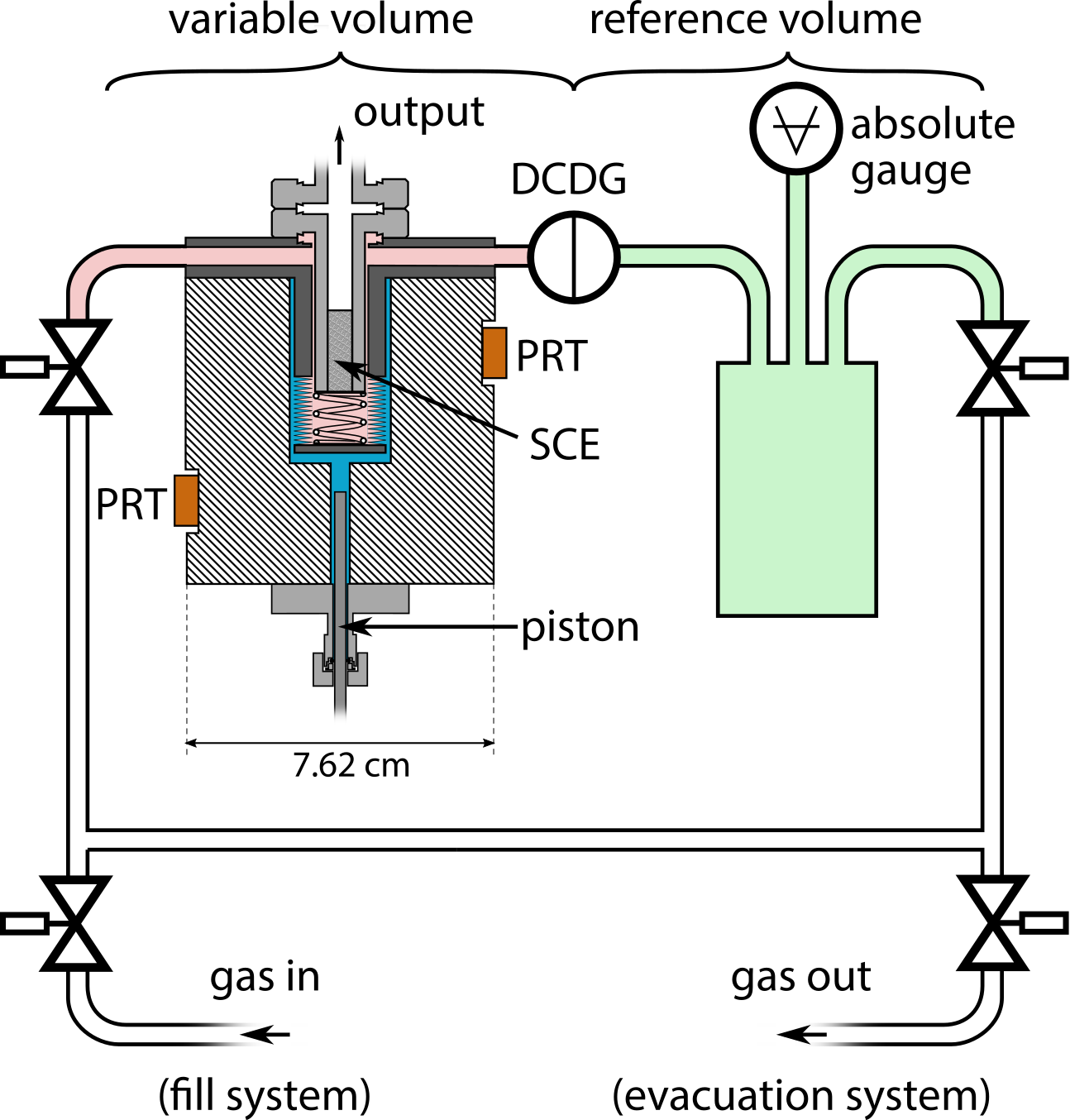}
    \caption{
    A simplified schematic of the flowmeter.
    The flowmeter is divided into two components: a variable volume (left) and reference volume (right).
    Pink denotes gas in the variable volume, green denotes gas in the reference volume, and blue denotes hydraulic fluid.
    Dark gray components are made of titanium; light gray components are made from stainless steel, and hashed components are made from aluminium.
    The following abbreviations are used: PRT: platinum resistance thermometer, SCE: standard conductance element, and DCDG: differential capacitance diaphram gauge.}
    \label{fig:schematic}
\end{figure}
\label{sec:components}
To meet the requirements outlined in Sec.~\ref{sec:intro}, we designed and built a new flowmeter, shown schematically in Fig.~\ref{fig:schematic}.
The flowmeter is roughly divided into two main components: a variable volume and a reference volume.
Two additional components, the fill system and evacuation system, will be described in a later publication; they are unimportant to the present discussion of the operation of the flowmeter and its uncertainty analysis.
Briefly, the fill system fills both the variable volume and reference volume with the same gas to a pressure within 2~\% of a target pressure in the range of 13~Pa to 150~kPa.
The evacuation system uses a combination of a roots pump to achieve low (rough) vacuum and a turbomolecular pump to achieve pressures $<10^{-6}$~Pa.
All valves are pneumatically actuated for fully automated operation, filling, and evacuation.
All uncertainties in this section and those that follow are standard uncertainties ($k=1$), unless stated otherwise.

Fig.~\ref{fig:schematic} shows the core components of the variable volume (bellows, leak, spring, oil canister, oil, and piston) to scale.
The entirety of the variable volume is approximately $13$~mL with a ``standard conductance element'' (SCE) providing the output.
The SCE is made from sintered stainless steel powder~\cite{Yoshida2012} with approximately $50$~nL/s conductance for N$_2$ near 300 K.
The variable volume is constructed from welded bellows made from titanium, which minimizes H$_2$ outgassing~\cite{Fedchak2021}.
All stainless steel components except the SCE were baked at $450$~$^\circ$C for at least 20 days in a vacuum furnace to minimize hydrogen outgassing~\cite{Fedchak2018}.

The bellows portion of the variable volume is surrounded by an aluminium canister that contains the hydraulic fluid (diffusion pump oil).
Prior to filling the canister, the oil is boiled for roughly 4 hours in a diffusion pump to remove any bubbles or dissolved gas.
To fill, the oil is forced from the diffusion pump and into the evacuated canister ($< 1$~Pa) by a back pressure of roughly 1.5~atm of Ar.
The oil flows first into the bottom of the canister, filling to the top in order to ensure that no trapped, residual gas could remain to form a bubble.
An additional 10~mL of oil is subsequently pushed through the canister to fill a reservoir attached to its top through a manual valve.
The pressure of the oil in the canister $p_{\rm oil}$ can be set with this reservoir by applying pressure on the surface of the oil in the reservoir from a regulated gas cylinder and opening the valve connecting to the oil canister and reservoir allowing them to come to equilibrium.
Afterwards, the valve is closed and the pressure $p_{\rm oil}$ is maintained in the canister.

Rather than measuring the pressure of the variable volume $p$ directly, we constantly compare $p$ against the pressure $p_{\rm ref}$ in a reference volume of approximately 200~mL.
This scheme has two advantages.
First, the differential pressure $\Delta p = p - p_{\rm ref}$ can be measured with a more precise CDG than the absolute pressure $p_{\rm ref}$, reducing noise in the feedback loop.
In the present case, a maximum range 133~Pa, differential, bakeable CDG measures $\Delta p$.
Second, a single differential CDG minimizes the volume of the variable volume compared to the multiple gauges that are needed to accurately measure the pressure over the full operating range of 13~Pa to 150~kPa.
Because the pressure rise due to displacement of the piston is proportional to the volume of the variable volume, minimizing the volume minimizes both the gain required and noise in the feedback loop.
The relative uncertainty in $u(\Delta p)/\Delta p = 1$~\% is dominated by uncertainty in the calibration of the differential CDG.

The choice of absolute pressure gauge(s) that measure $p_{\rm ref}$ is of paramount importance, as they are a dominant contributor to the uncertainty in $\dot{n}_{\rm out}$.
In this work, we generally use one transfer standard (made of three gauges) and one gauge to measure $p_{\rm ref}$; this setup is a typical configuration used for calibrations at NIST.
The transfer standard is based on three temperature-stabilized capacitance diaphragm gauges with maximum ranges 133~Pa, 1.33~kPa and 13.3~kPa~\cite{Ricker2017}.
This transfer standard is calibrated directly from mercury and oil ultra-sonic manometers and offers low uncertainty.
The uncertainty $u(p_{\rm ref})$ is a complicated function of $p_{\rm ref}$, but generally $u(p_{\rm ref})<2\times10^{-3} p_{\rm ref}$.
To extend to pressures above 13.3~kPa, we supplement this transfer standard with a resonant silicon gauge with maximum range of 130~kPa.
This RSG has been calibrated against a mercury manometer, and offers uncertainties of $u(p_{\rm ref})<1\times10^{-3} p_{\rm ref}$.
This combination is used in the data presented in Sec.~\ref{sec:flow_measurement}.

We also tested a suite of gauges chosen to minimize risk of contamination of the CAVS.
This suite consists of three bakeable capacitance diaphragm with maximum ranges 133~Pa, 1.33~kPa and 133~kPa~\cite{Scherschligt2021}.
Given its bakeability, it offers the lowest risk of contamination for XHV applications.
It generally has $u(p_{\rm ref})<3\times10^{-3} p_{\rm ref}$, except in the range between 1.33~kPa and 13.3~kPa, where $u(p_{\rm ref})$ can be as high as 1~\% of $p_{\rm ref}$.
This suite is used in Sec.~\ref{sec:comparison}.

The temperature of the variable volumes is measured using two industrial-quality platinum resistance thermometers (PRTs) mounted on the outside of the 7.62~cm diameter aluminium oil canister.
Approximately 2.5~cm of aluminium and 1.4~mm of oil separate the PRTs from the titanium variable volume, deteriorating the thermal contact between the gas inside the variable volume and the PRTs.
For this reason, we seal the flowmeter (reference volume, variable volume, and fill system) inside a temperature-controlled insulated box, and temperature stabilize the air inside to within 20~mK.
We estimate $u(T) = 36\mbox{ mK}$, with $30$~mK due to calibration uncertainty and $20$~mK being the typical measured difference between the two thermometers.
Self heating of the PRTs, measured to be about 3 mK, is negligible.

The insulated box doubles as an oven: heaters contained within the box can heat the flowmeter to 120~$^\circ$C to desorb any accumulated water from the interior surfaces.
All vacuum components, including all vacuum gauges, within this enclosure are bakeable under vacuum to 120$^\circ$C.
Bakes typically last 48~h, sufficiently long to leave hydrogen as the dominant outgassing component.
After baking, we estimate the base pressure of the flowmeter to be less than $10^{-6}$~Pa.

To compress the variable volume, a $10$~cm long, 3.1749(5)~mm diameter steel piston is inserted into the oil through a Viton o-ring seal.
Its depth is precisely controlled using a micrometer screw with a linear displacement rate of 0.499(1)~mm/turn.
The micrometer screw position is read using a digital rotary encoder with 2048 steps/turn, and we take the bit resolution as a $k=1$ uncertainty (a slight overestimate).
Combining the uncertainty of the dimensions and rotary encoder, we find that the uncertainty in the piston's displacement $\Delta V_{\rm piston}$ is given by $[u(\Delta V_{\rm piston})]^2=[1.4\times10^{-5}\mbox{ mL}]^2 + [(2.5\times10^{-6}\mbox{ mL})\times N_{\rm turns}]^2$, where $N_{\rm turns}$ is the number of turns measured by the rotary encoder.
For a typical $N_{\rm turns}>10$, $u(\Delta V_{\rm piston})/\Delta V_{\rm piston} \approx 6\times10^{-4}$.
The change in volume for the allowable travel of the piston corresponds to $\Delta V_{\rm piston} \leq 0.25$~mL.

The change of volume of the piston $\Delta V_{\rm piston}$ effects a change in the oil pressure $\Delta p_{\rm oil}$, which, in turn, changes the volume of the variable volume $\Delta V$ and the oil $\Delta V_{\rm oil}$.
The degree to which these volumes change is set by their respective $dV/dp$ values.
Because the volume of the oil canister is constant, the sum of their respective changes should sum to zero.
Thus,
\begin{equation}
    \Delta V_{\rm piston} =  -\left[\left(\frac{dV}{dp}\right)_{\rm VV} + \left(\frac{d V}{dp}\right)_{\rm oil}\right] \Delta p_{\rm oil},
\end{equation}
where $(dV/dp)_{\rm VV}$ and $(dV/dp)_{\rm oil}$ denote the $dV/dp$ values for the variable volume and oil, respectively.
Solving for $(\Delta V)$ yields
\begin{equation}
    \label{eq:displacement_correction}
    \Delta V = -\frac{\Delta V_{\rm piston}}{1 + \left(\frac{dV}{dp}\right)_{\rm oil}/\left(\frac{dV}{dp}\right)_{\rm VV}}\ .
\end{equation}
Clearly, to determine the degree to which $\Delta V = -\Delta V_{\rm piston}$ requires knowledge of both $dV/dp$.
For the oil, $\left(dV/dp\right)_{\rm oil} = -V_{\rm oil}/\beta$, where $V_{\rm oil}=8.2(8)$~mL is the volume of the oil in the canister measured via a gravimetric technique and $\beta$ is the oil's bulk modulus.
The oil is a four- or five-ring polyophenyl-ether [brand name Santovac 5(P)\footnote{Any mention of commercial products is for information only; it does not imply recommendation or endorsement by NIST nor does it imply that the products mentioned are necessarily the best available for the purpose.}] diffusion pump oil, for which $\beta\approx 2.4$~GPa is its measured bulk modulus from Ref.~\cite{Song1991} (at 37.8~$^\circ$C and 690~Pa).

\begin{figure}
    \centering
    \includegraphics{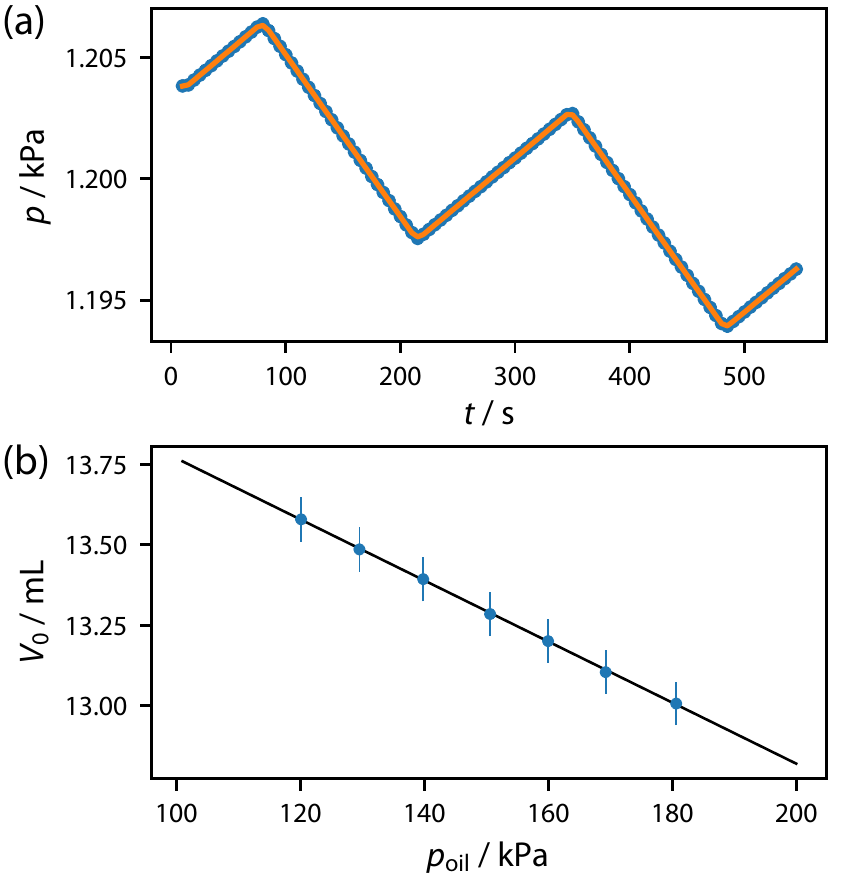}
    \caption{(a) Measurement of the dead volume $V_0$ of the variable volume.
    Total pressure in the variable volume $p$ is modulated with time $t$ by changing the volume with the piston.
    Blue points show data and orange curve shows fit.
    (b) Compressibilty of the variable volume.
    Extracted $V_0$ vs. oil pressure $p_{\rm oil}$ (blue points).
    The slope of a linear fit (blue line) determines $(dV/dp)_{\rm VV}$.
    The error bars represent the total uncertainty ($k=1$); the statistical uncertainty (type-A) is approximately the size of the points.}
    \label{fig:compressibility}
\end{figure}

The variable volume's spring constant and geometry set $(dV/dp)_{\rm VV}$.
When pressure is exerted on the bellows, assuming no bending of the walls of the bellows inward, the resulting net force will, by symmetry, be applied only to the bottom plate.
The resulting change in volume for an increase of oil  pressure can then be calculated as
\begin{equation}
    \left(\frac{dV}{dp}\right)_{\rm VV} = -\frac{1}{3} \frac{\pi A}{k}  \left( r_1^2+r_1 r_2+r_2^2 \right),
\end{equation}
where $A$ is the area of the bottom plate of the bellows, $r_1$ is the inner radius, $r_2$ is the outer radius, and $k$ is the spring constant.
By design, the spring constant of the bellows is approximately $9$~N/cm.
With such a small spring constant, evacuating the variable volume with the outside held at standard atmospheric conditions would cause the bellows to fully contract.
To prevent the bellows from contracting fully when evacuated, a spring with a spring constant of approximately $80$~N/cm is inserted into the variable volume to add rigidity.
After filling the canister with oil, the spring also keeps the oil pressure greater than atmosphere, thereby preventing an air bubble from leaking into the oil.
We estimate our spring constant $k\approx 90$~N/cm, resulting in $\left(\frac{dV}{dp}\right)_{\rm VV} \approx -1.2\times10^{-2}$~mL/kPa.\footnote{In the current NIST standard flowmeter, the bellows is prevented from fully contracting by hanging weights from it rather than inserting a spring.  This has the effect of providing a constant offset in the applied force, and does not affect its $dV/dp$, which is still set by the spring constant of the bellows alone.  We estimate its $\left(dV/dp \right)_{\rm VV} \approx -2.0$~mL/kPa}
Combined, we thus estimate $(dV/dp)_{\rm oil}/(dV/dp)_{\rm VV}\approx 4\times10^{-4}$, comparable to the relative uncertainty in the measurement of the piston's displacement.

We improve the estimate of $(dV/dp)_{\rm VV}$ by measuring the volume of the variable volume with the piston fully inserted $V_0$ (the ``dead'' volume) as a function of oil pressure.
$V_0$ is comprised of the plumbing connecting the input valve, leak, and differential CDG, the rigid portion of titanium chamber surrounding the SCE, some portion of the differential CDG itself, and the portion of the bellows volume that remains uncompressed when the piston is fully inserted (i.e., the pink regions in Fig.~\ref{fig:schematic}).
The volume is inferred by filling the variable volume to a pressure $p$ and modulating the variable volume using the piston.
Assuming an incompressible oil and an ideal gas at constant temperature $T$, the result is:
\begin{equation}
    p(t) = \frac{n(t) R T}{V_0 + \Delta V(t)}
\end{equation}
where $\Delta V(t)$ is the change in volume modulated by the piston, $n(t) = n_0 - \dot{n}_{\rm out}t + \dot{n}_{\rm OG}t$ is the number of moles at time $t$, and $n_0$ is the initial number of moles.
Figure~\ref{fig:compressibility}(a) shows an example of such a measurement, with a triangle wave modulation of the volume, which gives a best-fit value of 13.4(1)~mL for $V_0$, dominated by the uncertainty in $\Delta p$.
We note that the shift from assuming an incompressible oil is a factor of 5 smaller than the uncertainty.
By repeating this measurement at different oil pressures $p_{\rm oil}$ [Fig.~\ref{fig:compressibility}(b)], a linear fit extracts $\left(\frac{dV}{dp}\right)_{\rm VV} = -9.5(5)\times10^{-3}$~mL/kPa, within 25~\% of our crude estimate.

In a separate apparatus that uses the same oil canisters as the flowmeter, we measure the bulk modulus $\beta= 1.4(3)$~GPa.
It is unclear why our measurement of $\beta$ is lower than the literature value of 2.4~GPa in Ref.~\cite{Song1991}.
Our data cover a wide pressure range, and allow us to easily exclude a trapped bubble that would adversely impact the measurement.
It is more likely that the chemical properties of the diffusion pump oil we used are not identical to the oil tested in Ref.~\cite{Song1991}.
For the purposes of Eq.~\ref{eq:displacement_correction}, we use our reported value.
With these values, $\Delta V = -[1 - 5.9(1.2)\times10^{-4}] \Delta V_{\rm piston}$.

\section{Outgassing}
\begin{figure}
    \center
    \includegraphics{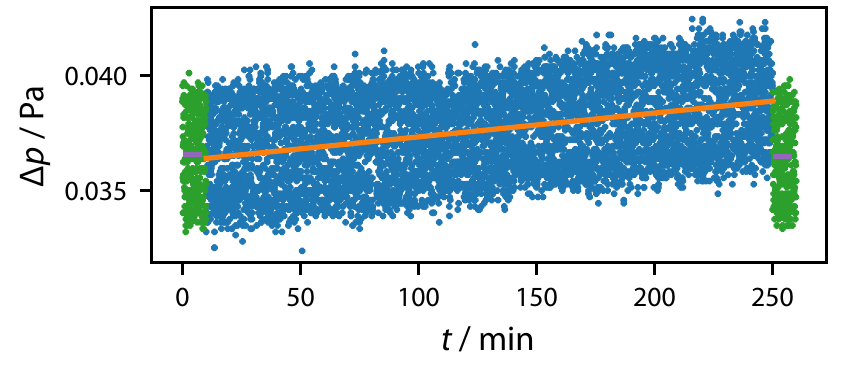}
    \caption{\label{fig:outgassing} Example 6~h outgassing measurement.  For ten minutes at the beginning and end, the offset in the differential CDG is measured (green points).  In the intervening 6~h, the variable volume is isolated and pressure rise measured (blue points).  The outgassing rate is determine by a linear fit of the pressure rise (orange), adjusted based on shift of the gauge offset (purple).}
\end{figure}

Outgassing, $\dot{n}_{\rm OG}$, can present a large systematic shift, particularly at low flows.
While inert gasses are typically used with the flowmeter, the contribution of additional gasses can be important depending on whether the measurement that uses the flow involves a gas-species-dependent pressure measurement (e.g. calibrating a spinning rotor gauge) or partial pressure measurement (e.g. calibrating a He leak).
In either case, outgassing will increase the pressure inside the variable volume, and because the differential gauge is equally sensitive to all gas species, cause a flow reported by the flowmeter that is smaller than that of the intended flow.

Because the present flowmeter is designed for low outgassing, we measure $\dot{n}_{\rm OG}<1\times10^{-15}$~mol/s under ideal conditions.\footnote{This is about a factor of 100 less than the current NIST standard flowmeter.}
At these low outgassing rates, the rate-of-rise measurement technique is still valid, but requires long times.
An example of a 6-hour outgassing measurement is shown in Fig.~\ref{fig:outgassing}, with a measured $\dot{n}_{\rm OG} = 9.6(9)\times10^{-16}$~mol/s.
During the entirety of this measurement, the reference volume is being continuously evacuated such that its pressure is constant and small.
For such measurements spanning several hours, we find that the differential gauge's zero offset (i.e., its reading at true zero differential pressure) can change by similar amounts as the actual pressure rise due to outgassing.
For this reason, we ordinarily take 10 minutes of data before isolating the variable volume and after reconnecting the variable volume to the pumping system to determine the drift in the gauge reading at zero pressure.
The dominant uncertainty in this measurement is statistical fluctuations; all other sources contribute less than $1$~\%.

The stated uncertainty of our outgassing measurement, dominated by statistics, is determined by the repeatability.
We typically observe a short-term repeatability (i.e., immediately repeated measurements) of $u(\dot{n}_{\rm OG})/\dot{n}_{\rm OG}\leq 10$~\% ($k=1$).
Long term repeatability is more variable: as the flowmeter is repeatedly filled and subsequently evacuated, the observed outgassing rate can increase up to values around $3\times10^{-15}$~mol/s, demanding another bake.
Thus, for daily operation of the flowmeter, we always first measure $\dot{n}_{\rm OG}$ before adding gas and measuring flows.
For these measurements of $\dot{n}_{\rm OG}$, we use the observed short-term repeatability of 10~\% as the $k=1$ uncertainty.

It is of interest to try to determine the source of the outgassing.
The treated (baked) stainless steel components and titanium comprise about 300~cm$^2$ of the variable volume.
Using a rough specific outgassing rate (i.e., per unit area) of $5\times10^{-12}$~Pa~L/s/cm$^2$~\cite{Fedchak2021}, these components should contribute roughly $6\times10^{-16}$~mol/s of outgassing.
One the other hand, the SCE has not, to the best of our knowledge, been baked to remove dissolved hydrogen.
Thus, the SCE's specific outgassing rate could be comparable with untreated stainless steel, which is up to 100 times larger than the other, treated stainless steel components~\cite{Fedchak2021}.
Even with this larger specific outgassing rate, the SCE's relatively smaller exposed surface area of roughly 20~cm$^2$ would limit its total outgassing contribution to roughly $4\times10^{-15}$~mol/s.
Thus, we suspect that the SCE is our dominant source of outgassing.
All of these estimates are within an order of magnitude of our observed outgassing rate.

\section{Flow measurement}
\label{sec:flow_measurement}
A typical flow measurement starts first by filling both the variable volume and reference volume to a target pressure to obtain, based on rough knowledge of the leak's conductance $C$, a target flow.
Subsequently, both the reference volume and variable volume are isolated by closing their respective valves.
A software proportional-integral-derivative (PID) controller then feeds back on the piston's displacement rate $\dot{V}_{\rm piston}=d\Delta V_{\rm piston}/dt$ to null the pressure difference, i.e., set $\Delta p =0$.
This configuration makes the PID output, $\dot{V}_{\rm piston}$, constant when the flow is stable, as $\dot{V}_{\rm piston}\approx -\dot{V} \approx -C$.
The valve actuation changes the pressure in both the variable volume and reference volume, causing them to rise slightly.
The variable volume, because it is smaller, rises more and thus $\Delta p >0$ initially.
Thus, the PID first retracts the piston to bring $\Delta p$ to zero and then settles in on its measurement value of $\dot{V}_{\rm piston}\approx -C$.
Once $\dot{V}_{\rm piston}$ reaches a constant value and $\Delta p=0$, the software is automatically triggered to start recording data.

\begin{figure}
    \center
    \includegraphics{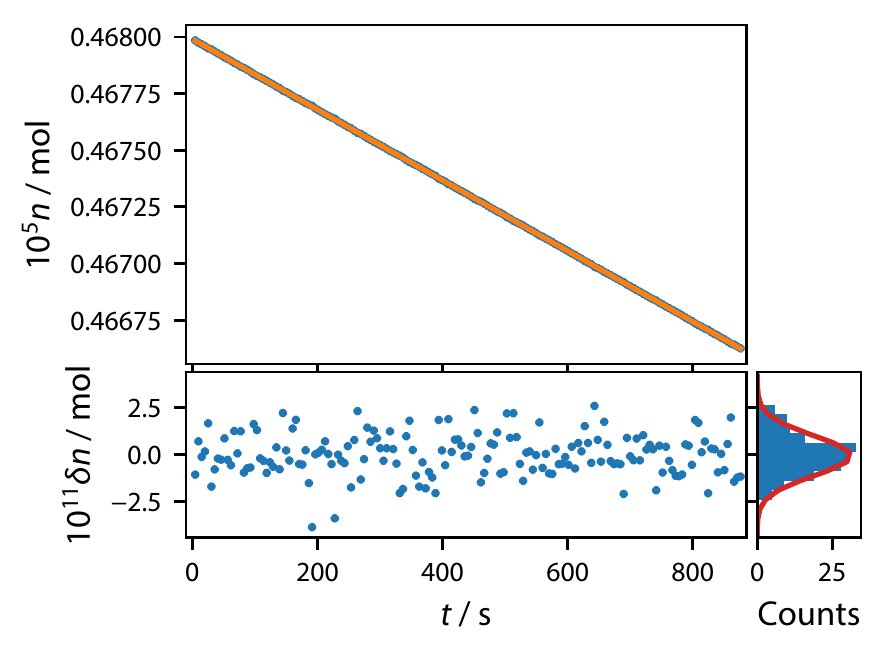}
    \caption{\label{fig:operation} Typical flow measurement using Ar at pressure $p=857.4(7)$~Pa and $T=296.78(4)$~K.
    (Top) The calculated number of moles in the variable volume $n$ vs. $t$.
    Blue points show data, orange line shows linear fit (see text).
    (Bottom left) Fit residuals $\delta n$ vs. $t$.
    (Bottom right) Binned fit residuals with Gaussian fit (red).}
\end{figure}

The reference volume absolute pressure $p_{\rm ref}$, the differential pressure $\Delta p$, the temperature $T$, and the piston's velocity and position are typically recorded every $\Delta t = 5$~s.
The length of a given run can vary anywhere between 300~s and 3600~s.
The volume $V(t)$ is then calculated from the piston position with the correction of Eq.~\ref{eq:displacement_correction},
and the pressure of the gas in the variable volume is computed through $p(t)=p_{\rm ref}(t)+\Delta p(t)$.
Even though $\Delta p=0$ under normal operation, we nevertheless record and use it in our calculation of $p(t)$; except at the very lowest of pressures, it is of no consequence.

To determine $\dot{n}$, we first use the computed $p$ and $V$ together with the recorded $T$ to solve the virial equation of state (Eq.~\ref{eq:virial_eos}) to determine the number of moles in the variable volume $n$ at all $t$.
Figure~\ref{fig:operation} shows an example run of the flowmeter at an Ar fill pressure of 857.4(7)~Pa and temperature $T=296.78(4)$~K, with a measured output flow of $\dot{n}_{\rm out} = 1.560(2)\times10^{-11}$~mol/s.
The computed $n(t)$ is shown as the blue points in the top panel of Fig.~\ref{fig:operation}; $n(t)$ decreases with time, as expect for all types of low-gas-flow vacuum flowmeters (see Sec.~\ref{sec:theory}).
The resulting $n(t)$ is subsequently fit to a line using standard least squares fitting.
This fit is shown as the orange line in the top panel of Fig.~\ref{fig:operation}.
The best fit slope $\dot{n}$, together with the outgassing correction, determine the output flow through $\dot{n}_{\rm out}=-\dot{n} + \dot{n}_{\rm OG}$.
This method simplifies the algebra of our measurement equation: we use Eq.~\ref{eq:virial_eos} to calculate $n(t)$ and fit for its derivative $\dot{n}$
rather than taking its derivative analytically first, obtaining Eq.~\ref{eq:full_measurement_equation_with_virial}, and subsequently computing $\dot{n}_V$, $\dot{n}_p$, $\dot{n}_T$, and their sum $\dot{n}=\dot{n}_V+\dot{n}_p+\dot{n}_T$
Moreover, this method accurately captures any correlations between $\dot{n}_V$, $\dot{n}_p$, and $\dot{n}_T$, consequently giving a more accurate determination of the type-A uncertainty of $\dot{n}$, denoted as $u_A(\dot{n})$.
The fit residuals $\delta n$ are shown in the bottom panel, and binned and fitted with a Gaussian in the bottom right.
The typical scale of $\delta n/n \approx 10^{-6}$ and the $\delta n/\Delta n\approx 10^{-3}$, where $\Delta n$ is the total change in $n$.
While $\delta n$ may appear Gaussian distributed, the residuals are in fact correlated, as we shall see.

As a consistency check, we also calculate the flow according to Eq.~\ref{eq:full_measurement_equation_with_virial}.
Specifically, the $V(t)$, $p(t)$, and $T(t)$ data are time-averaged to determine $p$, $V$, and $T$ and are linearly fit using least squares to determine $\dot{V}$, $\dot{p}$, and $\dot{T}$.
These results are combined to compute $\dot{n}_V$, $\dot{n}_p$, and $\dot{n}_T$ in Eqs.~\ref{eq:full_measurement_equation_with_virial}-\ref{eq:non_ideal_T_correction}, determining the flow.
The two techniques generally agree well within the statistical uncertainty of the fitted slopes.
This second technique can be sometimes overestimate the type-A  uncertainty of $\dot{n}$, particularly if any two components ($\dot{n}_p$, $\dot{n}_V$, or $\dot{n}_T$) have behavior that is of order $t^2$ or higher that cancel in the sum $\dot{n} = \dot{n}_p + \dot{n}_V + \dot{n}_T$.
Nevertheless, it does allow for determining each component's contribution to the flow.
For example, for the data shown in Fig.~\ref{fig:operation}, $\dot{n}_V/\dot{n}\approx 0.9939$ and $\dot{n}_p/\dot{n}\approx 0.0068$, and $\dot{n}_T/\dot{n} \approx -0.0009$.
Thus our flowmeter, while predominantly constant-pressure (i.e., driven mostly by $\dot{n}_V$), has some slight constant-volume behaviour $0.68$~\% and constant-volume-and-pressure behaviour $0.09$~\% as well.
Generally, measurements that show non-ideal behaviour where  $|\dot{n}_p/\dot{n}|>5$~\% and/or $|\dot{n}_T/\dot{n}|>0.3$~\% are discarded.
These values are chosen because they are roughly $5$ times larger than the measured standard deviation of the respective quantities, and ensure that the flowmeter is operating properly with no errors.

\begin{figure}
    \centering
    \includegraphics{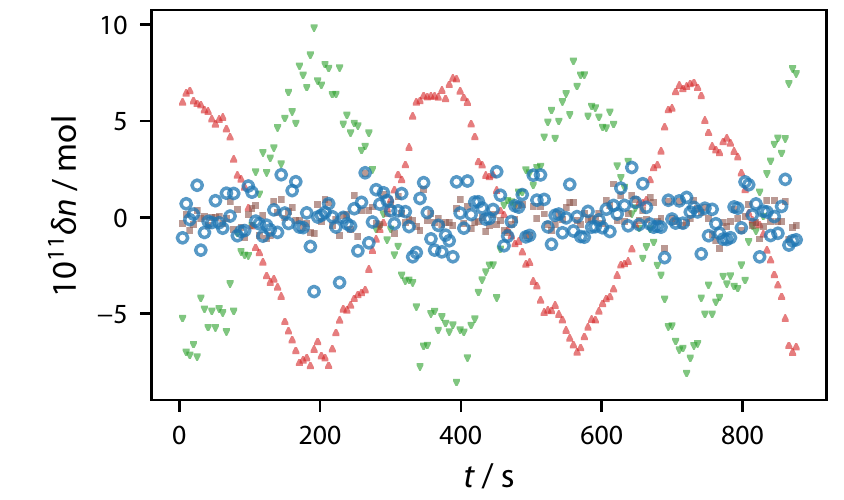}
    \caption{Correlations and components in the residuals of Fig.~\ref{fig:operation}.  The full residuals are shown as blue points, the red upward triangles show the component due to $\dot{n}_V$, green downward triangles show  $\dot{n}_p$, and brown squares $\dot{n}_T$.}
    \label{fig:residual_correl}
\end{figure}
The degree to which the three components of the flow are correlated can be illustrated more clearly by further analysis of the residuals $\delta n_i$.
Figure~\ref{fig:residual_correl} shows the contributions from the residuals of $\delta \dot{n}_V$, $\delta \dot{n}_p$, and $\delta \dot{n}_T$.
The scatter over short times is roughly equal for $\delta\dot{n}_p$ and $\delta\dot{n}_T$, whereas $\delta \dot{n}_V$ is the least noisy.
More interestingly, however, is the fact that there is a clear correlation between changes in $\delta \dot{n}_p$ and $\delta \dot{n}_V$.
We can clearly see that an oscillation in the absolute pressure $p$, which appears in $\delta\dot{n}_p$.
This oscillation is caused by a known instability in the laboratory's environmental control system.
The instability causes the plumbing components that connect the temperature-stabilized flowmeter to absolute pressure gauges to experience a temperature oscillation, thus affecting the pressure of the gas within.\footnote{These plumbing components are passively insulated; without insulation, the oscillation is about 20 times larger.}
Nevertheless, the PID, whose only input is $\Delta p$, responds and creates an equal and opposite correlation in $\delta \dot{n}_V$ as that seen in Fig.~\ref{fig:residual_correl}, such that the sum $\delta\dot{n}$ is constant.
In this example, the PID is compensating for pressure drifts $\delta p$ on the order of $\delta p/p\sim 10^{-5}$.

Because the components of the residuals are correlated, we must take care when extracting any type-A random uncertainty $u_A(\dot{n})$.
In practice, we take $u_A(\dot{n})$ to be the larger of the reported uncertainty from a single run, as shown in Fig.~\ref{fig:operation},  or the standard deviation of multiple, repeated identical measurements.

We now consider systematic, or type-B, sources of uncertainty.
For our nominally constant-pressure flowmeter, each term in Eq.~\ref{eq:full_measurement_equation_with_virial} contains contributions from several shared measurements.
Namely, every term includes a measurement of $p$, $T$, and change in time $\Delta t$.
The uncertainties in $p$ and $T$ were discussed in Sec.~\ref{sec:components}.
While the last two terms share a measurement of the {\it total} volume $V$, the first term measures $\Delta V$, which has roughly six times better (lower) relative uncertainty.
While the relative uncertainty in $V$ is larger [$u(V)/V\approx 0.005$], it is weighted by $(\dot{n}_p+\dot{n}_T)/\dot{n}_{\rm out}$, which must be less than $<0.05$ because of the exclusion criteria discussed above.
Thus, $u(V)$ does not represent a significant source of uncertainty.
With respect to the uncertainty in the elapsed time $\Delta t$, the timing is performed by the computer clock, which has been checked using network time protocol against NIST time and found to be accurate at the 50~ms level over a period of 10~h. It does not contribute meaningfully to the total uncertainty.

In addition to the uncertainty arising from the measured physical quantities, we also consider gas purity.
In our flowmeter, we use research grade, or 99.99~\% purity gases.
The way in which impurities affect the measurement depends on the application for which the flowmeter is used.
For example, if the flowmeter is used to calibrate He leaks, the detector is gas-species sensitive and so any impurities in the He gas will not contribute to the detector signal; however, they may contribute to $\Delta p$ and $p$, and thus can represent a $u_{\rm gas} = 0.01$~\% uncertainty.
Even for calibration of vacuum gauges, it is unclear how a gas impurity will affect the calibration.
If, say, the impurity is water, it will most likely be adsorbed onto stainless steel surfaces in the interior of the flowmeter, and once so trapped, it will fail to contribute uncertainty through either mechanism in the above example.
However, once adsorbed, it may subsequently desorb from the surface and contribute to outgassing.
Indeed, we see evidence of this process when we fill the flowmeter and subsequently evacuate it; adsorbed contaminant gases desorb and cause a power law pump out curve indicative of compounds like water.
Thus, it seems reasonable to always include the gas contribution as a relative uncertainty in the uncertainty budget.

The full uncertainty in $\dot{n}_{\rm out}$, derived from the measurement equation Eqs.~\ref{eq:full_measurement_equation_with_virial}-\ref{eq:non_ideal_T_correction}, is then
\begin{eqnarray}
    \left(\frac{u(\dot{n}_{\rm out})}{\dot{n}_{\rm out}}\right)^2 = & \left(\frac{u_A(\dot{n})}{\dot{n}_{\rm out}}\right)^2 + \left(\frac{u(p)}{p}\right)^2 + \left(\frac{u(T)}{T}\right)^2 + \left(\frac{u(\Delta t)}{\Delta t}\right)^2 \nonumber \\
    &  + \left(\frac{\dot{n}_V}{\dot{n}_{\rm out}} \frac{u(\Delta V)}{\Delta V}\right)^2 + \left(\frac{\dot{n}_p+\dot{n}_T}{\dot{n}_{\rm out}}\frac{u(V)}{V}\right)^2 + \left(\frac{u(\dot{n}_{\rm OG})}{\dot{n}_{\rm out}}\right)^2 \nonumber \\
    & + u_{\rm gas}^2,
\end{eqnarray}
where the values of $\dot{n}_V$, $\dot{n}_T$, and $\dot{n}_p$ are computing using the $\dot{V}$, $\dot{p}$, and $\dot{T}$ (extracted from the linear fits for $V(t)$, $T(t)$ and $p(t)$ described above) and their respective mean values.
Here, we have ignored contributions to the uncertainty from the non-ideal gas coefficients.
Their relative contributions scale as $c_i/(1+c_i) [u(p)/p]$ and $c_i/(1+c_i) [u(T)/T]$, and thus their relative contributions to $u(\dot{n}_{\rm out})$ are generally less than about $1.0 \times 10^{-5}$.

\begin{figure}
    \center
    \includegraphics{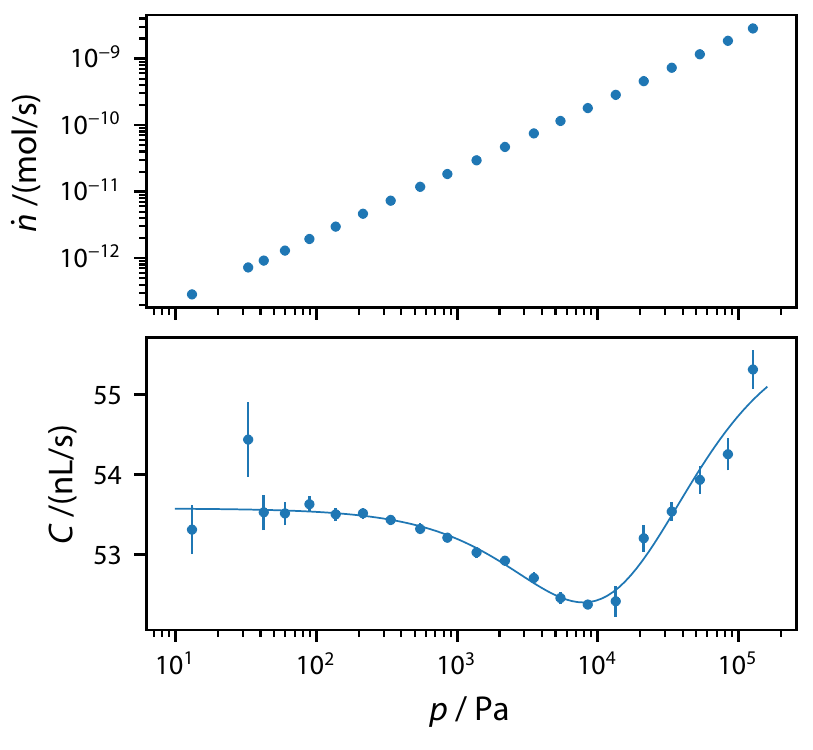}
    \caption{\label{fig:conductance and flow} Measured flow  of N$_2$ $\dot{n}$ out of the variable volume (top) and the associated conductance $C$ of the SCE (bottom) vs. fill pressure $p$.  Error bars are the $k=1$ uncorrelated combination of the type-A uncertainty, which is the standard error in the mean of four averages, and the type-B uncertainty (see text).  Curve in the bottom panel shows a fit to a ratio of polynomials in $P$.}
\end{figure}

As mentioned in Sec.~\ref{sec:intro}, the flow out is determined solely by the fill pressure $p$ and the conductance $C$ of the leak.
Figure~\ref{fig:conductance and flow} show $\dot{n}$ and $C$ as a function of $p$ for N$_2$.s
Because the SCE operates in the molecular flow regime to relatively high pressures, the relationship between $\dot{n}$ and $p$ is quite linear.
We fit the $C$ vs. $p$ data to a ratio of polynomials to extract for $C_0$ in the limit of $p\rightarrow 0$ and to help guide the eye.
The conductance shows a Knudsen minimum around $10^4$~Pa, which was not observed in Ref.~\cite{Yoshida2012}.
In the molecular flow regime ($p<100$~Pa) and assuming the temperature has negligible impact on the geometry of the SCE, the conductance $C$ does depend weakly on the temperature $C \propto \sqrt{T}$, so typical temperature fluctuations of less than $100$~mK should cause only $\delta C/C \approx 10^{-4}$.
Changing the gas species changes the conductance more dramatically as $C\propto 1/\sqrt{M}$, where $M$ is the molecular or atomic mass of the gas.
We only observe the expected scaling with mass roughly: we find that $C_0=53.55(4)$~nL/s, $44.76(5)$~nL/s, and $152.3(3)$~nL/s for N$_2$, Ar, and He, respectively.

\begin{figure}
    \centering
    \includegraphics{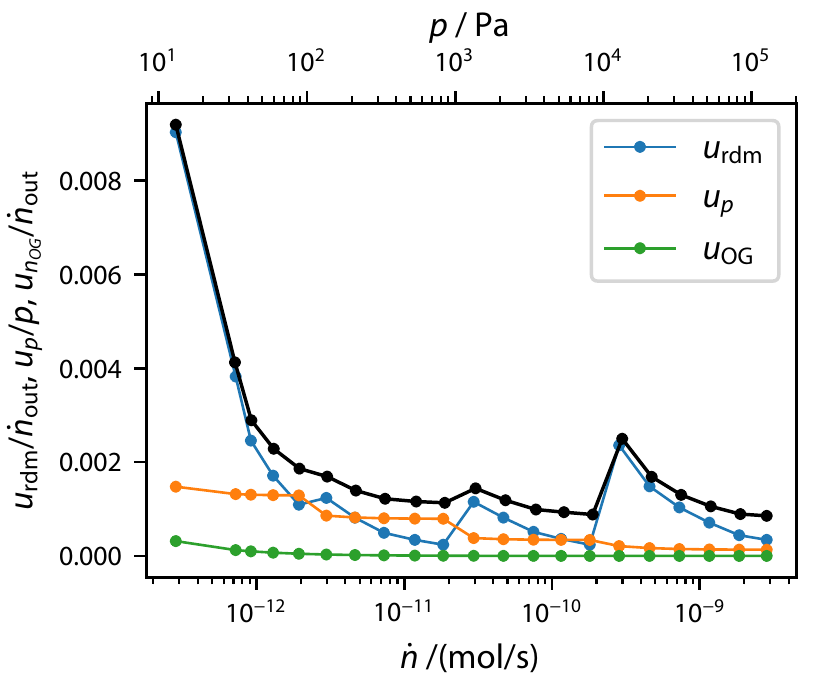}
    \caption{Uncertainty vs. flow for various components ($k=1$).  Black points represent the total uncertainty $u(\dot{n}_{\rm out})/\dot{n}_{\rm out}$.}
    \label{fig:uncertainty}
\end{figure}

\begin{table}
    \centering
    \begin{tabular}{l|c|c|c|}
         Source & $1.0 \times 10^{-13}$~mol/s & $1.0 \times 10^{-11}$~mol/s & $1.0 \times 10^{-9}$~mol/s \\
         \hline
         $u(p)/p$ & 0.0014 & 0.0008 & 0.0003 \\
         $[\dot{n}_V/\dot{n}_{\rm out}][u(\Delta V)/\Delta V]$ & $6.9\times10^{-4}$ & $6.9\times10^{-4}$ & $6.9\times10^{-4}$  \\
         $u(T)/T$ & $1\times10^{-4}$ & $1\times10^{-4}$ & $1\times10^{-4}$ \\
         $u(\dot{n}_{OG})/\dot{n}_{\rm out}$ & $9\times10^{-4}$ & $9\times10^{-6}$ & $9\times10^{-8}$ \\
         $[(\dot{n}_p+\dot{n}_T)/\dot{n}_{\rm out}][u(V)/V]$ & $<3\times10^{-4}$ & $<3\times10^{-4}$ & $<3\times10^{-4}$ \\
         $u(\Delta t)/\Delta t$ & $2\times10^{-6}$ & $2\times10^{-6}$ & $2\times10^{-6}$ \\
         $c_i$ & $<10^{-9}$ & $<10^{-7}$ & $<10^{-5}$ \\
         \hline
         $u_B(\dot{n}_{\rm out})/\dot{n}_{\rm out}$ & $0.0018$ & $0.0011$ & $0.0008$ \\
         \hline\hline
    \end{tabular}
    \caption{Type-B uncertainty budget at several different flows ($k=1$).  Calculations assume a displacement $\Delta V\approx 70$~$\mu$L.}
    \label{tab:uncertainty}
\end{table}

We can use the same data shown in Fig.~\ref{fig:conductance and flow} to examine how the values of $u_A(\dot{n})$ and $u(p)$ change dramatically with $p$ and $\dot{n}$.
Figure~\ref{fig:uncertainty} shows how these components change vs. flow rate.
Here, $u_A(\dot{n})$ is calculated using the residuals of a single fit.
Jumps in $u_A(\dot{n})$ and, to a lesser extent, $u(p)$ can be seen when the range changes in the absolute pressure gauge.
Outgassing begins to contribute below $1.0 \times 10^{-12}$~mol/s, where it reaches about 0.1~\% of the flowmeter's total uncertainty, and falls dramatically with higher flows.
Table~\ref{tab:uncertainty} summarizes the type-B contributions to the uncertainty in the flow $u_B(\dot{n}_{\rm out})$, showing the relative contributions of each term.

\section{Comparison}
\label{sec:comparison}
We have compared our new XHV flowmeter (XHVFM) to the existing vacuum flowmeter at NIST [known locally as the bellows flow meter (BFM)] used for spinning rotor gauge, ion gauge, and He leak calibrations over the overlap range of $1.0 \times 10^{-11}$~mol/s to $1.0 \times 10^{-8}$~mol/s.
To compare, He flows from both were both alternately routed to the new Vacuum Leak System (VALES) used for calibrating He leaks.
(This new system for calibrating leaks will be documented in a forthcoming article.)
In short, the VALES utilizes the chamber from the Primary Leak Standard (PLS) described in Ref.~\cite{Abbott1996}, in which a gas flow of He flows through an orifice and subsequently through a turbomolecular pump.
With this combination, a He flow of $10^{-12}$~mol/s will create a  partial pressure in the chamber of $10^{-8}$~Pa.

The partial pressure of He in the chamber is monitored using a quadrupole mass spectrometer (QMS, Hiden HAL 101-RC).
When monitoring the partial pressure $p_f$ generated by either flowmeter, the quadrupole data is accumulated for 5 minutes.
The first 2 minutes generally show transients that are related to switching the flow source; the last 3 minutes of data are averaged to obtain $p_{f, i}$.
For a given target flow rate, eleven partial pressures are recorded: five with the BFM $p_{{\rm B}, i}$, indexed from $i=1$ to $i=5$, filled to pressures that nominally produce 0.8, 0.9, 1.0, 1.1 and 1.2 times the target flow produced.
Six time-bracketed partial pressures generated by flow from the XHVFM, nominally producing the target flow, are recorded, $p_{{\rm X},i}$, with $i=1$ to $i=6$.
For each trace, the respective flowmeter records data to measure its own flow.

Quadrupole residual gas analyzers are known to drift and be non-linear~\cite{Becker2014}.
To correct for drifts, we form a partial pressure ratio by averaging the bracketed XHVFM data, $r_i=(p_{{\rm X}, i} + p_{{\rm X}, i+1})/2p_{{\rm B},i}$.
This technique compensates for the drift in the QMS at first order.
The flow from the the XHVFM according to the BFM can then be computed through $\dot{n}_{{\rm X-B},i} = r_i \dot{n}_{\rm B}$, where $\dot{n}_{\rm B}$ is the reported flow from the BFM.
For flows where $\dot{n}\lesssim 10^{-10}$~mol/s, we find that $\dot{n}_{{\rm X-B},i}$ is independent of the partial pressure ratio $r_i$.
In this case, we take the mean, denoted by $\langle \dot{n}_{\rm X-B}\rangle$, of the nominally five reported $\dot{n}_{{\rm X-B},i}$, and use the standard deviation as an estimate of the statistical uncertainty in the partial pressure measurement.
For larger flows, $\dot{n}\gtrsim 10^{-10}$~mol/s, $\dot{n}_{{\rm X-B},i}$ is at least a linear function of $r_i$, indicating non-linearity in the QMS.
When this occurs, we fit the $\dot{n}_{{\rm X-B},i}$ vs. $r_i$ to a linear or, at most, quadratic polynomial and evaluate it at $r = 1$ to determine the mean $\langle \dot{n}_{\rm X-B}\rangle$.
In this case, we evaluate the uncertainty in $\langle \dot{n}_{\rm X-B}\rangle$ by assuming $\chi_\nu^2=1$ for the fit and evaluating the $1$-$\sigma$ prediction interval at $r=1$.

\begin{figure}
    \centering
    \includegraphics{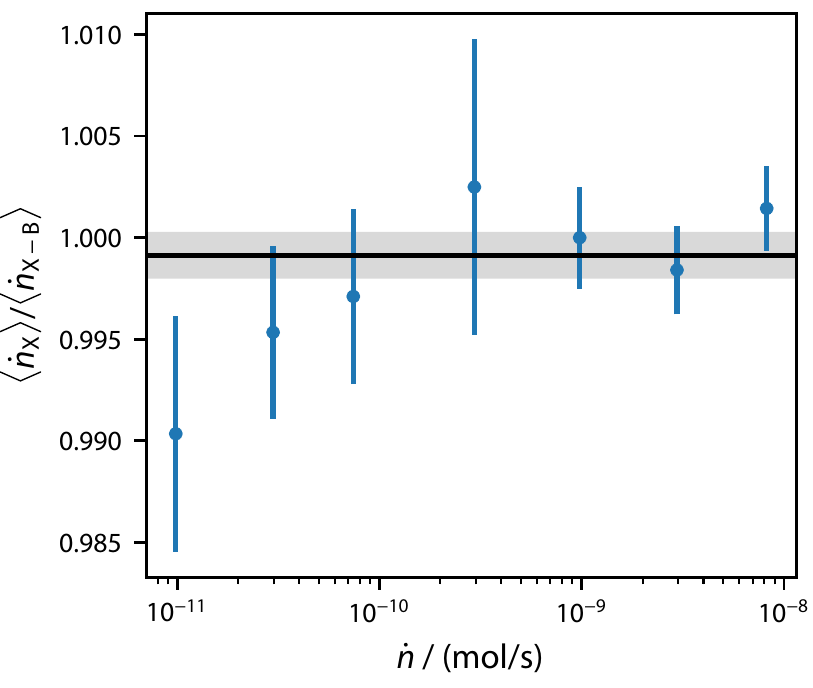}
    \caption{Ratio of the XHV flowmeter's reported He flow $\langle \dot{n}_{\rm X}\rangle$ to the same flow measured by the current NIST standard flowmeter (BFM) and vacuum leak system (VALES), $\langle \dot{n}_{\rm X-B}\rangle$.
    Error bars show the uncorrelated combination of type-A and type-B $k=1$ uncertainties of both flowmeters and the VALES.
    Solid horizontal line shows the weighted mean of the points; the gray band shows its $k=1$ uncertainty.}
    \label{fig:comparison}
\end{figure}

Figure~\ref{fig:comparison} shows the ratio of XHVFM reported flow, $\langle \dot{n}_{{\rm X}}\rangle$, to the XHVFM's flow as measured by comparison to the BFM, $\langle \dot{n}_{{\rm X-B}}\rangle$.
There are three sources of uncertainty that contribute to the error bars of Fig.~\ref{fig:comparison}: the uncertainty in the XHVFM measured flow, the uncertainty in the BFM measured flow, and the uncertainty in the VALES.
For most of the data, the type-A uncertainty in the VALES system is the largest contribution to the total uncertainty.
At $\dot{n}=2.95(2)\times 10^{-9}$~mol/s, the type-A uncertainty of the XHVFM dominates, because at this flow we are forced to use an absolute pressure gauge at high amplification, resulting in more random noise.
At $\dot{n}=9.83(4)\times10^{-12}$~mol/s, both the type-A uncertainty of the VALES and the type-A uncertainty of the XHVFM contribute roughly equally.
Given that type-A uncertainties are dominant for much of the data, we expect and find that 2/3 of the points encompass the mean within their $k=1$ uncertainty.

The two flowmeters agree to within their mutual uncertainty.
A weighted average of the points in Fig.~\ref{fig:comparison} indicates the XHVFM reports a flow 0.09~\% lower, but with a $k=1$ uncertainty of 0.12~\%, with a reduced $\chi^2_\nu$ of 0.83.
An unweighted average indicates a difference in flow of 0.21~\%, with a $k=1$ uncertainty of 0.14~\%.
In both cases, $\langle \dot{n}_{{\rm X}}\rangle/\langle \dot{n}_{{\rm X-B}}\rangle=1$ is captured at the $k=2$ level.
Using the latter, the flowmeters agree within $0.5$~\% with 95~\% ($k=2$) confidence.

\section{Conclusion}
We have described the operation and uncertainty of a new, hydraulic, constant-pressure flowmeter capable of operating down to $2.0 \times 10^{-13}$~mol/s.
In order to achieve accurate measurement at such low flows, we made two major improvements.
First, we reduced the outgassing in the variable volume to less than $10^{-15}$~mol/s.
To reduce hydrogen outgassing, our flowmeter uses titanium vacuum parts where possible, and vacuum-fired stainless steel parts elsewhere.
To reduce water outgassing, the entire flowmeter is bakeable to 120~$^\circ$C.
Second, we reduced the conductance $C$ of the leak which generates the flow.
By utilizing a standard conductance element (SCE)~\cite{Yoshida2012}, we achieve a conductance of about 50~nL/s for N$_2$.
Because the rate of volume change is nominally set by the conductance of leak, the flowmeter uses smaller pistons and micrometer screws, but these components are dimensioned to the same relative accuracy as our previous flowmeter to minimize overall uncertainty.
Our new XHV flowmeter has a type-B relative uncertainty $<0.2$~\% ($k=1$) over its entire range.

At the higher fill pressures needed to generate flows $\dot{n}_{\rm out}>1.0 \times 10^{-9}$~mol/s, we have quantified non-ideal gas corrections in terms of the virial equation of state.
These flow corrections can be non-negligible compared to the reported uncertainty for the present flowmeter, and for other similar flowmeters.
They affect not just constant-pressure flowmeters, but also constant-volume flowmeters.

Finally, we have compared our flowmeter to the NIST standard flowmeter currently used for calibrating ion gauges, spinning rotor gauges, and He leaks.
The two agree within their reported uncertainties at the $k=1$ level.
Specifically, we have found agreement within 0.5~\% at 95~\% confidence.
Currently, it is being used to measure thermalized cross sections in the cold atom vacuum standard (CAVS)~\cite{Scherschligt2017, Eckel2018}.
In the future, we anticipate it being used for calibrations of leaks and vacuum gauges.

\section*{Acknowledgments}
We thank Jacob Ricker and Jay Hendricks for useful discussions regarding CDGs, Patrick Egan for useful discussions regarding virial coefficients, John Stoup for dimensioning our pistons and micrometers, and Robert Berg and Patrick Abbot for a careful reading of the manuscript.

\section*{References}
\providecommand{\newblock}{}

\end{document}